# E-Semiotics

## Semiotics, Knowledge Management, and New Information Technologies

**Prof. DDr. Peter Stockinger**

Institut National des Langues et Civilisations Orientales (INALCO)
Fondation Maison des Sciences de l'Homme
Paris - France

**Conference**
**Oy FountainPark Ltd - Renewal Forum**
**(Helsinki, November 23rd, 2001)**

Peter Stockinger
Institut National des Langues et Civilisations Orientales (INALCO)
Fondation Maison des Sciences de l'Homme (FMSH)
54, Bd. Raspail – 75006 Paris
email : stockinger@msh-paris.fr
site web : http://www.semionet.fr

**Contents**

Peter Stockinger
Institut National des Langues et Civilisations Orientales (INALCO)
Fondation Maison des Sciences de l'Homme (FMSH)
54, Bd. Raspail – 75006 Paris
email : stockinger@msh-paris.fr
site web : http://www.semionet.fr

## 1) Plan of this paper

In this communication, we shall present a specific conceptual approach in the conception, development, and management of new (interactive) information and knowledge products and services. This approach is e-semiotics (like "electronic" semiotics, i.e., semiotics for the building, understanding, and managing of digital information and knowledge products and services).

E-semiotics is a rather new term. So, I will start this paper with a global definition of what e-semiotics is.

In the following chapter, I will compare e-semiotics with "traditional" semiotics (this means, with semiotics applied to the description and understanding of not-digital information products and services such as newspapers, professional communication, social activities, and so on). I will show that e-semiotics is only a special but new and rather technical domain of expertise for semiotic research and applications.

After, I will develop a clear understanding of the place of e-semiotics in knowledge management. I will show that knowledge management can be understood from different points of view. Two of them are of a particular importance for our purpose: the conceptual approach (i.e., the specification of models, templates, scenarios, scripts, etc. of information products and services) and the technological approach for the development of concrete products and services, their maintenance, re-use, etc.

I will dedicate an important chapter to concrete application domains where e-semiotics is used/will be used: multi-support electronic publishing; digital libraries, ontologies, semantic websites and information

Peter Stockinger
Institut National des Langues et Civilisations Orientales (INALCO)
Fondation Maison des Sciences de l'Homme (FMSH)
54, Bd. Raspail – 75006 Paris
email : stockinger@msh-paris.fr
site web : http://www.semionet.fr

agents, intranets, portals, and organizational memories, dynamic websites and web services.

In a last chapter, I will give some general information about possible benefits of e-semiotics for social organizations, understood especially in the sense of information and knowledge producing, sharing, and consuming entities.

Peter Stockinger
Institut National des Langues et Civilisations Orientales (INALCO)
Fondation Maison des Sciences de l'Homme (FMSH)
54, Bd. Raspail – 75006 Paris
email : stockinger@msh-paris.fr
site web : http://www.semionet.fr

## 2) What is e-semiotics, roughly speaking ?

Presented in a nutshell, e-semiotics stands for a set of methods, services, and tools for the conception and specification - the "scenarization", the "scenario building and management" - of new information services and products:

- stand-alone products and services as well as web-based, networked ones such as websites, distributed and heterogeneous information systems, and web portals
- in all different fields of e-activities (e-learning, e-culture, e-content, e-publishing, e-commerce, etc., …)
- and for all "traditional" sectors of activities.

As we will see again, there are two central notions here:

The *scenario* or *script,* which is - almost like in the film production - a kind of model, a "norm" which guides the development, implementation of information products and services as well as their evaluation in terms of efficiency, usability, understandability, and so on.

The *information project* which - like in other project oriented activities - knows several specific phases - from the user needs analysis until the release and delivery of a concrete service or product, via the conceptual and technical specification of this product or service, its technical realization, its implementation in a working place, etc. [cf. [STO 99]).

As already mentioned, a scenario is a kind of "model" a "guide", or a "quality norm" for the different activities concerned with the definition, production, release, maintenance, etc. of a product or service.

Peter Stockinger
Institut National des Langues et Civilisations Orientales (INALCO)
Fondation Maison des Sciences de l'Homme (FMSH)
54, Bd. Raspail – 75006 Paris
email : stockinger@msh-paris.fr
site web : http://www.semionet.fr

As we will see again, scenarios and models you need even for the most basic information systems, such as simple informational websites, product catalogs, document catalogs, and so on. Their need is much more obvious for the production and understanding of complex interactive information products or services such as corporate portals, e-learning systems, or, again, information agents.

The following graphical schema summarizes the three principal roles of a semiotic or structural scenario in an information and communication project of a social organization broadly speaking and of companies, more particularly:

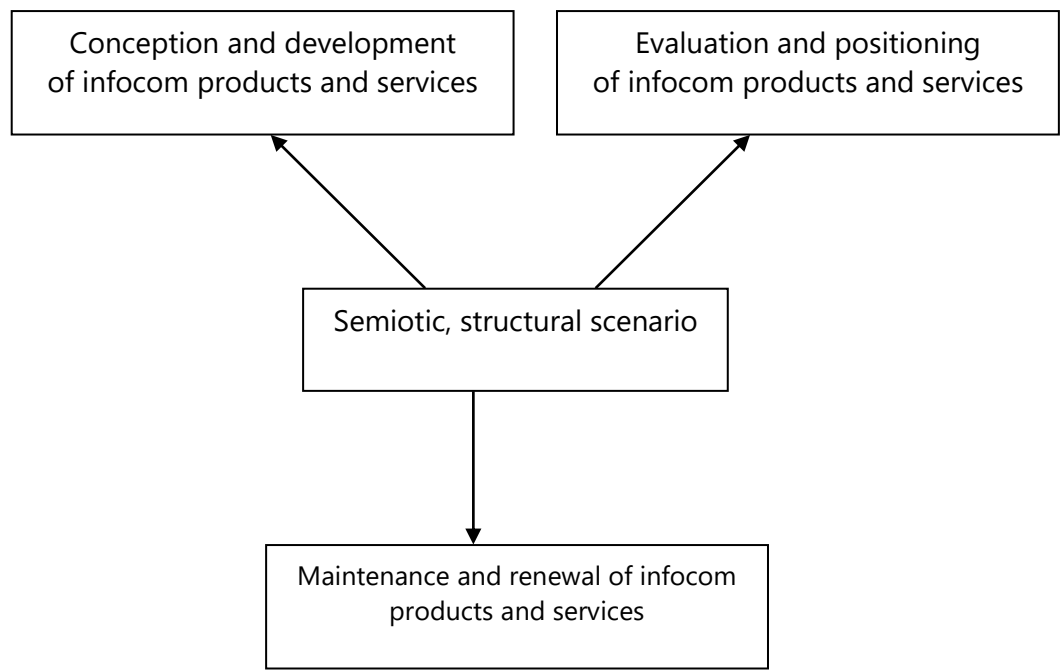

Peter Stockinger
Institut National des Langues et Civilisations Orientales (INALCO)
Fondation Maison des Sciences de l'Homme (FMSH)
54, Bd. Raspail – 75006 Paris
email : stockinger@msh-paris.fr
site web : http://www.semionet.fr

But a scenario or a script can also present itself as a - what is sometimes called - *coded description*, a *coded scenario*. A coded description or a coded scenario is a (more or less complex) semiotic description interpreted in the language (the syntax) of a specific technological standard or again in the environment of a specific software tool with the help of which a concrete information service or product will be developed.

For instance, in document management, an important issue is to provide richer models of the contents of the technical, commercial, juridical … documentation of a company. This means - generally speaking - that:

- On the one hand, a content description is produced, based on, and motivated by a structural theory of the contents of textual signs.

- and on the other hand the output of such a description - i.e., a content scenario - will be re-interpreted, for instance, in conformity with the syntax of XML or Dublin Core to produce "computer readable", reusable (standardized) models which can be exploited by any XML-compliant document management system.

This is an important issue because it constitutes one of the most central technological, institutional, cultural and economic challenges we already face, and we will face much more again in the coming years.

These issues are labeled under the head of *"web services"* - a kind of technology massively developed actually by Microsoft under its Microsoft.net platform, by its principal competitors like Sun, but also - as Open-Source initiatives - by academic institutions like the Stanford University and MIT.

A web service is something what an - individual or collective user (a company, a university, an administration, etc.) needs for being able to

Peter Stockinger
Institut National des Langues et Civilisations Orientales (INALCO)
Fondation Maison des Sciences de l'Homme (FMSH)
54, Bd. Raspail – 75006 Paris
email : stockinger@msh-paris.fr
site web : http://www.semionet.fr

benefit (more or less easily) of all the potentialities of the new information technologies : personal websites, corporate websites, virtual libraries, e-commerce applications, etc.

The point is that these web services are not really conceived as stand-alone applications but as services for which the user is subscribing an exploitation license in becoming a "member" of the big community of the web service provider (such as, for instance, Microsoft).

A (coded) semiotic description or scenario can be interpreted as a (more or less) specialized web service that can be used for the development and the management of a class of concrete products or services.

For instance, simple semiotic scenarios of corporate websites for small and very small enterprises include:

- a company presentation template,
- a product catalog template,
- a document directory model,
- a business news model,
- a business directory,
- etc.

Such scenarios can be coded, at least in simplified versions as *HTML - XML* and *ASP templates*, as well as in the form of a *relational database service*. They constitute a kind of library of services that can be achieved with new ones, versions of the existing ones, and so on.

Obviously, such models or templates can be more or less sophisticated, can possess a more or less user-adaptable graphical or visual interface and can provide more or less user attuned services.

Peter Stockinger
Institut National des Langues et Civilisations Orientales (INALCO)
Fondation Maison des Sciences de l'Homme (FMSH)
54, Bd. Raspail – 75006 Paris
email : stockinger@msh-paris.fr
site web : http://www.semionet.fr

One of the advantages of such a library is to allow small and very small entities so exploit the new information technologies for their information and communication needs without internal technical or technological competencies.

This approach that consists in the conceptual or semiotic specification and development of an (open and modular) library of web services becomes, indeed, more and more popular in a high diversity of concrete application domains:

- in e-commerce,
- in e-culture (virtual exhibitions, museums, …),
- in e-learning,
- in e-tourism,
- in e-community (social and community services, …),
- etc.

They constitute, without any doubt, the central piece of any development in the field of information and communication portals. A portal, typically, is built and maintained around "dynamic" information and communication components that are instantiated by the above-mentioned web services.

We will come back to this issue.

Peter Stockinger
Institut National des Langues et Civilisations Orientales (INALCO)
Fondation Maison des Sciences de l'Homme (FMSH)
54, Bd. Raspail – 75006 Paris
email : stockinger@msh-paris.fr
site web : http://www.semionet.fr

### 3) What are the differences to "traditional" semiotics?

This exposé is not conceived as a theoretical contribution to semiotics in general and to a specific semiotic trend in particular.

So, very generally and also loosely speaking, semiotics is understood, here, as a set of theories and methodologies for

- the description of information-loaded objects called signs,
- their use by any kind of cognitive agents (individuals, institutions, artificial agents, ...)
- as well as their history (their evolution).

Typically, information-loaded objects are documents of any kind (texts, pictures, graphics, audiovisual material, etc.) by the means of which a social organization produces, consumes, stores, and distributes information and knowledge : about itself, about its activities, about its competitors, about its missions and objectives, etc.

Other kinds of information-loaded objects are what are called "structured objects": databases, SGML or XML-compliant files, and specific computer programs such as information agents.

But information-loaded objects also include social activities and interactions (professional activities, daily life behavior, ...) that manifest themselves in a more or less stereotyped way, artifacts of any kind, or again, the different forms of a social space constraining people to behave regarding given - explicit or traditional - rules.

These are three major groups of "information loaded" objects that constitute the domain of expertise in semiotic description and evaluation, which are guided by a general theory and methodology of structural organization of signs and their use.

Peter Stockinger
Institut National des Langues et Civilisations Orientales (INALCO)
Fondation Maison des Sciences de l'Homme (FMSH)
54, Bd. Raspail – 75006 Paris
email : stockinger@msh-paris.fr
site web : http://www.semionet.fr

In this sense, really significant differences are not so much between "traditional" semiotics and e-semiotics than between different approaches and doctrines within the semiotic research world - most similar to any other scientific discipline.

> E-semiotics means especially: methodologies, services and tools for the production and understanding of *digital media* information products and services. In this sense - e-semiotics is the choice of a particular object, domain of reference, and that's all.

Naturally, there are at least superficially differences between digital media products and services and those realized on traditional media. Such differences are:

- the interactivity
- the hypertextuality
- the capacity to produce personalized versions of one and of the same product
- the reusability of components of products and services
- the distributedness of a product or service (i.e., the content of one product can be physically distributed over different places)
- Etc (cf. [STO 99]).

Nevertheless, these differences and others have sometimes been overestimated or exaggerated. Many scholars, especially in the literary studies, have shown that, at least structurally, the quoted phenomena can also be observed in "traditional" information products - maybe in a much more reduced, much more elementary form.

But there are some more specifically technical requirements in e-semiotics. This means that, basically, any person specifying or defining a product or service scenario has to know at least globally the technological constraints. This is a common-sense requirement. In general, people who do this work are people with a "double competence.

> - a technical and
> - a conceptual, semiotic one.

For instance, the specification of scenarios for interactive tourist guides are done by people who at least have a general idea of authoring systems and software tools, of CD-ROM technology, and so on.  People who have to specify a scenario of a distributed, heterogeneous information portal of a company have at least general notions of particular servers, database and document management technologies, etc.

This requirement becomes an almost necessary condition if semiotic descriptions or scenarios have to be coded, conformed to technological standards, or integrated in the workspace of given specialized software systems.

Note: Ideally speaking, these two competencies should be achieved by a third one: a competence in the field, the domain of expertise in which one has to work. For instance, to develop an interactive, encyclopedic product in the political history of a country requires an at least global familiarity not only with the historical "facts" but especially with the different conceptions, theories, and interpretations of these "facts" (an issue of which the importance is always underestimated).

Finally, e-semiotics as such is already a much differentiated domain which knows plenty of specializations. I want to quote here only some of them without any claim to be exhaustive.

There is the project of a computer semiotics or a computational semiotics, initiated principally by Peter Borg Anderson. Based on the sign theory of Peirce, this project consists, for instance, in a sign oriented description of computer applications (software, languages …).

Organizational semiotics aims at the understanding and modeling of information-loaded (i.e., "meaningful") signs within social organizations and is used for the sign-oriented description and management of complex information systems. The advantage of this approach regarding other methodologies for building and managing information systems is, without any doubt, its systematic and fine-grained approach to the meaning of cognitive resources used and produced in an organization for its information needs.

Narrative theory and narrative semiotics are used in the constitution of what is called "organizational memories. This means in the collection and processing of stories, by the means of which people of an organization (a company, for instance) verbalize their - professional - experiences. Without the techniques of story telling and the processing of produced stories, sometimes highly valuable but not formalized knowledge would get lost.

A popular and already well-known approach in e-semiotics is the use of structural semantics (meaning configurations and their constituents) for the elicitation and description of knowledge in textual resources (cf. [VOG 89]; [RAS 87]). In combination with conceptual graph theory ([SOW 84], [FAR 91], [STO 93]), structural methods for the identification and elicitation of "concepts" (i.e., notions) are used in the fields of knowledge representation and artificial intelligence.

Document semiotics is specialized in the analysis and description of documents of all sorts in the different fields of the information industry (written press, television, book industry, cinema, interactive digital media, etc.). The point here is that in document semiotics, a document is not understood in the traditional sense but much more as a family, a "community" of services that interact to satisfy the information needs or interests of users, of different user groups. In this sense, a tourist guide, for

Peter Stockinger
Institut National des Langues et Civilisations Orientales (INALCO)
Fondation Maison des Sciences de l'Homme (FMSH)
54, Bd. Raspail – 75006 Paris
email : stockinger@msh-paris.fr
site web : http://www.semionet.fr

instance, not only incorporates an "information service" of a specific country or region (a guide provides via text and images information of a country or a region) but also a series of other services: a voyage-preparation service, practical directory services, a practical hints and ads service, a bibliographical service, and so on. The advantage of such a conception is that it allows drawing parallels between forms or genres of traditional information resources and digital, interactive ones. It shows that between a traditional, paper-based document (such as a tourist guide) and an interactive website or web portal (for tourism), there exists much more of a continuum in terms of services than a radical opposition in terms of media support ([STO 94]; [TUR 97], [STO 99]).

There are many other such examples that would show that e-semiotics is already a much diversified, highly active domain of research and development.

Peter Stockinger
Institut National des Langues et Civilisations Orientales (INALCO)
Fondation Maison des Sciences de l'Homme (FMSH)
54, Bd. Raspail – 75006 Paris
email : stockinger@msh-paris.fr
site web : http://www.semionet.fr

## 4) E-Semiotics and knowledge management

Let me precise only quickly the actual general context that determines not only the business life but most of all parts of the social life in general (such as the administration, the political life, the education, the daily life, etc.).

This general context can be characterized by a whole series of notions that are, actually, very fashionable: "knowledge turn", "cognitive turn", "information economy", "knowledge economy", "virtual community", «virtual enterprise », «enhanced enterprise», and so on.

Even if we should deal with some critical distance with such notions and similar ones, they show nevertheless three points that are important for us:

1) As it has already been stressed, information and knowledge possess, for most of our institutions and social organizations, a strategic, functional, and commercial value.
2) The gathering, classification, circulation, exploitation, storage, and transfer of information and knowledge have therefore been recognized as a set of the most central, most important activities for more or less all social entities in our society.
3) Even if the emphasis on information and knowledge as central social and symbolic goods, as a "capital" in the sense of the French sociologist Bourdieu, cannot be seen in a direct dependency to the emergence of the NIT, the NIT, at least, accelerate this evolution which should be considered to be an irreversible one and which, progressively, will lead to new forms of social organizations as well as to new economic models.

These three points stick to the very complex challenges to which is confronted knowledge and information management.

Peter Stockinger
Institut National des Langues et Civilisations Orientales (INALCO)
Fondation Maison des Sciences de l'Homme (FMSH)
54, Bd. Raspail – 75006 Paris
email : stockinger@msh-paris.fr
site web : http://www.semionet.fr

*Knowledge management* means - in a nutshell - the lead and coordination of all activities in a social organization that have to do with the building, implementation, exploitation, and maintenance (evolution) of its information and communication products and services. It can be divided into different aspects or "approaches" among which the following three are the central ones:

Peter Stockinger
Institut National des Langues et Civilisations Orientales (INALCO)
Fondation Maison des Sciences de l'Homme (FMSH)
54, Bd. Raspail – 75006 Paris
email : stockinger@msh-paris.fr
site web : http://www.semionet.fr

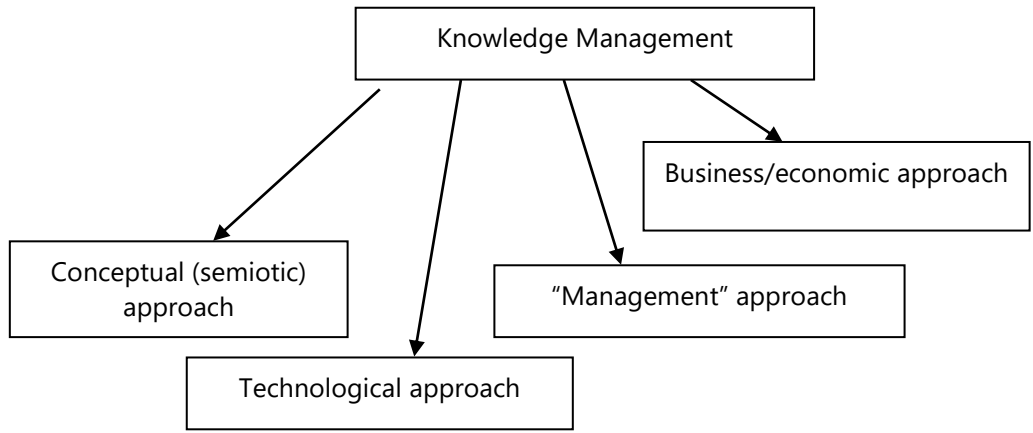

It is not our intention to develop here in a more detailed way all these aspects of knowledge management. We are interested more particularly in the semiotic or conceptual aspect as well as in the technological one.

Nevertheless, let us stress only that the *managerial approach in knowledge management* has to be aware of highly critical, not-technical questions such as the degree of:

- acceptance of an information system by its users,
- its easy-to-use by its users
- its easy-to-integrate in an existing working environment
- etc.

Another issue in the managerial approach in knowledge management is the definition and implementation of workflow diagrams that are adapted to the unavoidable technical constraints of an information or knowledge system, such as, for example:

- The identification of responsibilities  (of accountabilities)
- The (re-)definition  of activities and tasks regarding the role and the place of an information and knowledge system in an institution

- The training of people who have to deal with an information system
- And so on.

All these issues are, as it is known, highly critical and - psycho-sociologically speaking - complex ones. For instance, it is evident to everyone that the implementation and use of new information services privileges specific competencies and professional profiles to the disadvantage of other ones, etc.

As far as the *economic approach of knowledge management* is concerned, let us quote only the really difficult task of specifying realistic economic models for conceptual and technological innovations.

Let me refer here to a discussion among publishers who are confronted with the whole potentiality of electronic publishing. Electronic publishing is indeed a technological revolution. It also allows the definition and implementation of entirely new structures and forms in publishing: "New structures" in the sense that the content providers themselves can become their publishers and distributors; "New structures" also in the sense that an "electronic publisher" can be integrated, without any big technical problems, within a digital library application or again a virtual university application.

But these new structures are even less dramatic than the changes concerning the products themselves. Electronic publishing, together with the new content structuring technologies and standards (XML, Dublin Core, MPEG 7, etc.), allows the personalization of products, the dynamic updating of contents, the versioning of a same content, the multi-support publishing (i.e., via the web, on a CD-ROM, as a paper product, etc.).

For all these conceptual and technological opportunities, viable, adequate, and realistic economic models have to be defined. This is more than difficult, especially for small, specialized publishers as well as for

Peter Stockinger
Institut National des Langues et Civilisations Orientales (INALCO)
Fondation Maison des Sciences de l'Homme (FMSH)
54, Bd. Raspail – 75006 Paris
email : stockinger@msh-paris.fr
site web : http://www.semionet.fr

general publishers. This leads to the paradox that, on the one hand, most of the publishers agree that the NIT will change radically their profession and that they have to adapt their business to this new context. But on the other hand, most of them also confess not to know how to adapt, how to integrate these potentialities in their ongoing business, and how to prepare for the necessary technological, conceptual, and socio-economic changes.

In the next chapter, I will shortly discuss the conceptual (semiotic) and the technological aspects of knowledge management.

Peter Stockinger
Institut National des Langues et Civilisations Orientales (INALCO)
Fondation Maison des Sciences de l'Homme (FMSH)
54, Bd. Raspail – 75006 Paris
email : stockinger@msh-paris.fr
site web : http://www.semionet.fr

## 5) The semiotic and technological approach in knowledge management

As already mentioned, it is not our intention in this paper to discuss in a more detailed way theoretical or methodological issues in semiotic research. I only would like to enumerate some basic assumptions in e-semiotics, and once more come back to the relationship between semiotic scenarios and web services.

### 5.1) The semiotic approach in knowledge management.

In a few words, it can be understood regarding three basic and intuitively clear questions:

*First question:* What is knowledge? Is there any difference between knowledge and information, and if yes, what is the difference ?

Knowledge, intuitively speaking, is a "cognitive resource" that someone needs for satisfying a goal, completing a task, behaving appropriately in a given context, interpreting signals and messages in a "right" way, etc. It is also called "*intelligence*" or "strategy" for acting, deciding, judging…

But it is also a *value*: there are, regarding a given norm, preferable, better knowledge, more valuable knowledge, etc. There is also false knowledge, negative knowledge, etc.

In semiotics, knowledge is similar to *language* - language not only for being able to understand and produce information but also for interacting with people of a community, with the social world, etc.

So, *in describing knowledge or knowledge structures, you describe essentially - from a semiotic perspective - language or languages*: technical languages, juridical languages, the languages of fashion, the languages of administration, and so on.

A language contributes to what is called a culture, a system of knowledge and values you need to assimilate if you want to work, interact, life within a given institution or community.

Information, more precisely speaking, is the set of the signals of a situation that you can interpret and understand with your knowledge. Much of the words that are for me not understandable may constitute significant signals - information - for somebody else who possesses such a language, such a culture.

*Second question:* where do we find knowledge and information?

Knowledge and information we find in signs, in any kind of object that vehicles or is supposed to vehicle some information for a human or artificial agent (the user, the reader, the spectator, the listener). There are classes of objects - artifacts - specialized in doing this job: texts, images, films, drawings… These classes of objects are organized, as known, in "archives," "libraries", "documentation centers", "information centers", and so on.

But, virtually, all kinds of objects can become an "information-loaded sign" and therefore vehicle information for some cognitive agent. Let us think only of the interpretative work that has to be done in the historical sciences: paleontology, geology, climatology, and so on. All these sciences have to "read" natural objects to understand the evolution of a geological formation, a species, or the climate. In this sense, natural objects can become "information-loaded signs" for groups of users.

The dealing with "information loaded" signs to achieve some pre-defined purposes within a social organization is called an *information*

*project* based on an *"information contract"*. Such pre-defined purposes are, for instance, to constitute a corporate information broker, a corporate archive, a corporate learning system, and so on. Their realization is the object of the "information contract".

An information project knows several key activities, among which the three following ones are more particularly connected with the dealing of "information-loaded" signs:

- Information watch (i.e., the localization of "information-loaded signs", of resources containing relevant information for a user or a social organization)

- Information organization (i.e., the description, indexing, and annotation of "information loaded signs" in a database or a file directory system)

- Information exploitation (i.e., communication, sharing, exchange, collaborative work, … on stored, accessible "information loaded" signs and the production - authoring and editing - of information or knowledge products and services)

*Third question:* how to describe information loaded signs in order to access to the relevant information - to its content -, to be able to store a description of this information, to reuse, share, communicate and exploit it for publishing or other objectives.

This third question concerns more particularly the theoretical and methodological issues of semiotics in general and text or discourse semiotics in particular.

Peter Stockinger
Institut National des Langues et Civilisations Orientales (INALCO)
Fondation Maison des Sciences de l'Homme (FMSH)
54, Bd. Raspail – 75006 Paris
email : stockinger@msh-paris.fr
site web : http://www.semionet.fr

Without developing this crucial question in a more detailed way, let us say only that there are different scopes of a semiotic description of a sign (a class of signs):

- The structural composition of a sign or a class of signs, i.e., the underlying organizational patterns of the contents of a sign and the expression of this content;
- The production process (the "genesis") of a sign or a class of sign (covering typical activities and phases);
- The interpretation and, more particularly, exploitation context of a sign or a class of signs
- The context of management of "information-loaded" signs or classes of signs.

Without developing this essentially theoretical and methodological point, let us quote the principal general description criteria on which a semiotic description, the building of a semiotic scenario is based (for further information, see [STO 99] and the on-line documents on the semionet.com website[1])

*1) Corpus criteria*
- documents, objects, activities constituting the corpus
- test corpus
- validation corpus

*2) Structural criteria concerning the organization of an "information loaded object" (a set of "information-loaded objects")*
- content criteria
- expression of content criteria
- text, media support criteria

---

[1] Cf. http://www.semionet.com

Peter Stockinger
Institut National des Langues et Civilisations Orientales (INALCO)
Fondation Maison des Sciences de l'Homme (FMSH)
54, Bd. Raspail – 75006 Paris
email : stockinger@msh-paris.fr
site web : http://www.semionet.fr

*3) contextual criteria (of use/exploitation of an information-loaded object, a set of "information-loaded objects").*
- objectives and uses
- institutional settings

*4) Descriptive criteria (to be applied for the description, the scenario itself)*
- level of "granularity",
- level of exhaustiveness,
- level of technicality.

### 5.2) The technological approach of knowledge management

This approach has to do especially with the finding of technological solutions for developing and maintaining interactive information products or web services. Technological questions are concerned especially with:

- Hardware and software solutions for developing concrete information or knowledge products and services (such as websites, digital libraries, portals, but also interactive handbooks, e-learning tools, and so on).
- Standards in, for instance, information processing and exchange, document management, text processing, image and video processing, etc.

Standards have to ensure different important issues in information technologies such as:

- the readability of concrete information or knowledge products and services such as the above quoted ones by people working with different platforms or software tools;
- the re-usability of pieces, components of such products and services within other ones;

- the common access and the sharing of these products and services
- the possibility to evolve them in towards other services or products,
- etc.

There exists a significant number of international or national standards. One of the actually most prominent standards is XML (the extensible markup language) for a better structuring of the content of information-loaded signs. But one has to be aware:

- These standards only propose a common syntax, a common "formalism".
- There may exist as many descriptions as users of a collection of knowledge objects.
-

As already indicated in the second chapter of this paper, there is a "new" piece of technology that is becoming more and more important - the *web services*.

It is certainly one of the most important research and development domains, given - as it seems - the high economic and commercial impacts of this technology for its "owners".

The general idea "behind" web services is to offer a user (an individual or a social organization, no matter) for subscription a series of basic or more or less specialized "tools" that allow him to do what he/she needs/wants to do on the web: building a personal website, building a corporate website, creating a news list, realizing a virtual exhibition, constituting a personal archive, authoring an on-line course in semiotics, and so on.

There are several strategic issues to which we would like to stick:

Peter Stockinger
Institut National des Langues et Civilisations Orientales (INALCO)
Fondation Maison des Sciences de l'Homme (FMSH)
54, Bd. Raspail – 75006 Paris
email : stockinger@msh-paris.fr
site web : http://www.semionet.fr

- The web services (the "programs") remain on a host server, on the server of the organization that is the "owner" of them.
- The user can exploit them (download a version of them, work remotely, etc.) after having registered himself (this is the new "passport" system of Microsoft) and paid a subscription.

- The user can use the space and capacities of the host server (as this is the case, for instance, for the "pre-web" service Hotmail of Microsoft.

- In respecting the technological and juridical constraints, new web services can be developed by third parties, who may host their services on their servers.

- A distinction has to be made between basic web services (Microsoft, for instance, has identified basic services such as MyContacts, MyCalendar, MyCoordinates, ...) and more or less specialized ones (such as, for instance, web services specialized for the building of corporate websites, websites specialized for building and managing a web-based content archive, etc.)

It is not the issue of this paper to discuss possible and even probable consequences of this technology. But it appears to be compelling, even if there is a high probability that it will be confronted with serious obstacles.

As we will see in the next chapter, the place of e-semiotics in this field of technology concerns, especially, the specification and modeling of specialized web services - an exiting but highly complex research and development domain.

Peter Stockinger
Institut National des Langues et Civilisations Orientales (INALCO)
Fondation Maison des Sciences de l'Homme (FMSH)
54, Bd. Raspail – 75006 Paris
email : stockinger@msh-paris.fr
site web : http://www.semionet.fr

## 6) Fields of concrete applications

As we have already seen before, e-semiotics is relevant for a high variety of conceptual and mixed - conceptual, technological and sociological - questions in the field of new information technology and especially knowledge management. We want to discuss here very shortly the - so to speak - interest - of e-semiotics within the following six application fields :

- Multi-support and, personalized publishing,
- Digital libraries,
- Ontologies, semantic web, and information agents
- Dynamic website production and management
- Intranets, corporate portals, and organizational memories
- Modular and distributed web services.

### 6.1) Multi-support and personalized publishing

This field of research and development is one of the most important and central ones, especially for the

- production (authoring),
- description,
- indexing,
- user-filtered distribution,
- permanent maintenance.

… of specialized and very specialized information. Such specialized and very specialized information is produced, for instance, in public or private research labs, but also by professionals for professionals. It is highly valuable information, but under very specific conditions, such as:

Peter Stockinger
Institut National des Langues et Civilisations Orientales (INALCO)
Fondation Maison des Sciences de l'Homme (FMSH)
54, Bd. Raspail – 75006 Paris
email : stockinger@msh-paris.fr
site web : http://www.semionet.fr

- It is really relevant only for a defined variety of goals.
- It is really "understandable" only by a few people.
- It is rather perishable but can also, contrarily, contain knowledge of which the value is recognized only after a more or less important time span.
- Its intrinsic evaluation is very problematic - no real methods exist for that.
- It is stored, in general, with a lot of other (specialized) information in more or less voluminous information data (books, handbooks, articles, CD-ROMs, etc.).

Technically speaking, a (web based) service for the multi-support and personalized publishing of specialized information is composed of at least by the following components:

- A library of "contents" (i.e., of information data like text files, video files, images, animations, structured data, etc. - information data in which store and vehicle specialized information);

- Models ("templates"), i.e., more or less sophisticated scenarios or scripts by the means of which a) the contents are indexed and retrieved and b) organized in a way that corresponds to defined information genres (such as "minutes", "articles", "thematic folders", ...)

- Tools for editing contents in predefined templates

- Tools for multi-support publishing of selected and edited contents (as websites, CD-ROMs, paper-based documents, etc.).

E-semiotics is concerned here with two strategic issues:

- the indexing, description and also annotation (commentary production) of contents;
- the specification of information genres

Concerning the first issue - the indexing, description, and annotation of content - it has to be stressed out that a given piece of information (a piece of "content", such as, for instance, a business event or a political event, is processed on different levels :

- The thematic level (i.e., type of the event, actors, and roles implied, social and historical context, events path, etc.)
- The rhetoric or narrative level (how one "speaks" about a given event : in the form of a simple description, in giving an explanation, in developing a discussion around it, in the form of examples and testimonies, etc.)
- The discourse level (who "speaks" about a given event ?, to whom ?, in applying for the opinion of other actors ?, in focalizing on which items ?, …?)
- The level of the expression modalities of a piece of content (the piece of content is expressed in a textual form? in a visual form? what is the overall draft or lay out, the storyboard?)

Concerning the second point, the specification of a semiotic scenario corresponding to an information genre, has to consider the following aspects:

- The functional units composing a scenario (i.e., thematic sequences and scenes in which pieces of information are developed),
- The different accesses to specific information,
- The different types of exploration ("navigation", "(interactive) reading") of information developed in a several sequences or scenes,
- The different types of aids for a correct exploitation of an information product or service,

Peter Stockinger
Institut National des Langues et Civilisations Orientales (INALCO)
Fondation Maison des Sciences de l'Homme (FMSH)
54, Bd. Raspail – 75006 Paris
email : stockinger@msh-paris.fr
site web : http://www.semionet.fr

- The different types of the visualization of information contained and developed in an information product or service (cf. also [STO 99]).

### 6.2) Digital libraries

In speaking about digital libraries, one has to be aware that a digital library is not necessarily restricted to being a kind of copy of the organization and services proposed by the "traditional" institution called a library. In fact, digital libraries may evolve (and they are doing it) towards more complex and also new forms of services such as:

- Multi-support publishing,
- E-learning,
- Portals for thematically restricted virtual communities,
- etc.

We would like to quote here an important French R&D project called Opales (Outils pour des Portails Audiovisuels Educatifs et Scientifiques) and which is headed by the Institut National de l'Audiovisuel (INA). One of the main objectives of this project is to develop a set of services, methods, and aids for the filming, digitalization, description, and indexing, as well as communication of conferences, seminars, symposia, etc. via the web. Such events are daily life events for research organizations and contribute massively to the constitution of a scientific memory as well as to the kind of "corporate culture" of a given scientific organization.

But the project Opales is not only concerned with the specification and development of technical services, methods and aids for building and managing digital libraries. Another central concern of it is to enable researchers, teachers, but also other professional users (journalists, authors, etc.) to "process" digital files containing a conference or a seminar. This means, concretely speaking, that a professional user should be able:

Peter Stockinger
Institut National des Langues et Civilisations Orientales (INALCO)
Fondation Maison des Sciences de l'Homme (FMSH)
54, Bd. Raspail – 75006 Paris
email : stockinger@msh-paris.fr
site web : http://www.semionet.fr

- To identify and delimit those sequences in a digital video (as well as those chapters or paragraphs in a textual document) that are of interest for him.
- To index, describe, and comment on them, following his needs.
- To establish his working archives that he could share with those with whom he has to work together.

Coming back to e-semiotics, its concern here is to deliver a model, a scenario of how people describe, "read", or "interpret" an audiovisual document. This is a necessary to enable a user to do - within a given model - his descriptions and comments and to organize them following nevertheless a commonly shared standard or norm.

Another contribution of e-semiotics in this context consists in the elaboration of thematic grids, or "graphs" helping to organize a thesaurus, to ensure his follow-up and to attune it to specific user needs and requirements.

This second contribution leads already to the problem of (semantic) ontologies we will speak about shortly in the next paragraph. As known, there are some very prominent and widely used "meta-standards," or again, generic models, for defining the thematic configuration that is typical to a given domain of expertise. Such generic models concern, especially, the classification of items or data in a domain of expertise - taxonomy or mereology (part-whole classifications). But there exists, naturally, a huge variety of other such possible generic models for describing, for instance, actions and plans, goal-oriented behavior, concurrent and polemic situations, etc.

Which types of generic models are adequate for describing sets of conferences or seminars depends, naturally, on the contents of these events. Conferences can have, for instance:

- A highly historical (chronological, biographical) content,

- A theoretical-explanatory content,
- A theoretical-critical content ("comments"),
- A practical application oriented content,
- A classificational content,
- Etc.

The specification of scenarios or scripts representing these (and other) forms of (*rhetorical* or - generally speaking - *narrative*) content processing - can then be used for placing at the disposal of professional users (teachers, researchers, journalists, …) a highly sophisticated environment for describing, commenting, reusing, sharing, …. contents produced in conferences and other scientific events.

### 6.3) Ontologies, semantic web and information agents

As it is well known, one of the most central challenges, actually, is the identification and elicitation of not only relevant but even more value-added information or knowledge. This is, more particularly, true for the web and the incredible quantity of available information. For instance, the result of a search of relevant information on the web with the help of the engine Google concerning the item "corporate culture", is of about 161 000 possible answers, the result of a search concerning the item "knowledge management" is of about 590 000 possible answers, etc.

Important research efforts are today dedicated to what are called "ontologies," i.e., knowledge representations of given domains of expertise. Besides more specific ontologies, there also exist more general and even very general ones that are based on simple definitions of the lexicon of a language like English to give a user the possibility to specify the meaning of a word that he uses in his search of relevant information. For instance, the word "Paris" can have the meaning "capital of France", "town in Texas", "figure in the Greek mythology", etc. There exist already tools (prototypes) that use this strategy of de-ambiguizing the meaning of a word.

For instance, the (composed) word "knowledge management" could be at least decomposed into the four meanings we have defined above: technological meaning, conceptual or semiotic one, managerial one, and business one.

Nevertheless, we feel that there are many problems with such a "semantic rich" approach: how to justify the choice for a set of meanings, how to be sure that these definitions are shared by everyone, how to handle the problem of the evolution of meanings, how to consider concurrent meaning structures, etc.

Moreover, in taking again our example of "knowledge management", there are many other criteria that could be considered for limiting the scope of a concrete search. For instance:

- There can be a big difference if "knowledge management" is localized on a web page or if it covers a whole website (i.e., if it is localized, hypothetically, in the title of the homepage of a website).

- A website dedicated to "knowledge management" can be a .com website, a research website, a pedagogical one, and so on. Information and "orientation" of information depend at least partially on the general purposes of a website.

- More particularly, the word "knowledge management" may only be *used* as a general or specified notion in one web resource, but in another one it may be *introduced explicitly* by the means of a definition or a description. In the first case, one has to interpret the use of the word, in the second case, the author already gives the interpretation.
-
- In one web resource, there may be only one - "authoritative" - definition or description of "knowledge management" whereas in another web resource lists of definitions and descriptions,

Peter Stockinger
Institut National des Langues et Civilisations Orientales (INALCO)
Fondation Maison des Sciences de l'Homme (FMSH)
54, Bd. Raspail – 75006 Paris
email : stockinger@msh-paris.fr
site web : http://www.semionet.fr

discussions of several concurrent definitions or descriptions, or again examples of one or more definitions may be found. We see here once more again that efficient information or knowledge management has to consider what is called the *rhetorical* or - in a general sense - *narrative schemes* underlying the development of a piece of information in an information data. In one web resource, the notion of "knowledge management" may play a central role (but it may be designed linguistically by a more or less important number of lexical expressions), it may constitute a "director theme" among all the themes that constitute the specific content of an information data. In another web resource, it may constitute only a rather peripheral role, a marginal one.

Obviously, this list of criteria is only an ad hoc and illustrative one. It shows, nevertheless, one - from a semiotic perspective - important aspect of a single web resource in particular and of the web in general: both behave like *information loaded signs*, like more or less complex discourse activities within a community of discourses (cf.[STO 01]) about specific "subjects" or "topics" -

- That may be of more or less interest for a given user,
- In which a given user wants to enter to become a member of this community,
- For which a given user has to show a specific competence, a specific identity, etc.

Information agents are small, more or less autonomous programs that can be configured for specific information needs. For instance, there exist information agents that search the web (or parts of the web) for relevant business information or again for relevant financial information. To enable such programs to do that job, one has to specify the themes (in general: sets of alphanumerical characters) in which one is interested. Other specifications may concern the period and delay of an information search, the geographical localization (which is highly problematic) of the search,

Peter Stockinger
Institut National des Langues et Civilisations Orientales (INALCO)
Fondation Maison des Sciences de l'Homme (FMSH)
54, Bd. Raspail – 75006 Paris
email : stockinger@msh-paris.fr
site web : http://www.semionet.fr

the exclusion of information written in particular natural languages, and so on.

Different research is already undertaken to enhance the "understanding" of organizational features of a content piece by an information agent (i.e., the descriptive model of a domain of expertise that an information agent has "inside" him). It appears to be obvious that more complex, more sophisticated descriptive models will consider features such as those we have quoted above and for which we need a semiotic understanding of web resources in terms of discourse, discourse activity and discourse community.

Finally, one of the most appealing actual researches considers societies of agents, collective agents that have to cooperate to achieve a specific goal, such as the search of relevant information for some users. This idea of agents specialized in specific activities and tasks is not only an exciting one but corresponds to our sociological intuition of workflow in social organizations based on the notion of a "labor division." In any case, working with a set of agents presupposes the description, the scripting of the interactions between agents, the sharing of common knowledge and values, the evaluating of actions, and so on. Such complex action patterns are known as narrative schemas in the structural theory of Greimas [[GRE 76], [GRE 79]).

### 6.4) From dynamic websites to web services

A lot of research and development activities are actually concerned with the building, management and evolution of organizational intranets. Such Intranets are to be considered as the digital mode of communication and information production, communication, exchange and sharing between groups and individuals of a social organization, the social organization itself and other ones (cf. the case of the B-to-B communication), the social organization and their clients or users in the most general sense as well as the social organization and its - so to speak - social "Umwelt" (press and media, political and administrative context, ...).

In one of our European sponsored R&DT project - A Virtual Hypermedia Factory - coordinated by Olivetti, one of the central purposes has been to develop tools and methods enabling small and very small institutional structures as well as professional users to build and maintain their websites without knowing much more than the use of Explorer or Netscape.

In analyzing, for instance, the global organizational communication and information schema of small companies (like in the fields of arts and crafts) or of public structures (like research labs, teaching institutes, libraries, and archives), it is possible to decompose such a schema in some recurrent structural and functional units. Each of these units, as it can be shown, possesses a more or less stereotypical organizational model. Such a model can be used as an input for defining and building rather simple database services. Combined with a simple web-based management interface (for inserting, updating, and deleting structured information as well as for transferring not-structured information - texts, pictures, videos, animations, ... - to a web server) a professional user can build without any technical knowledge a corporate website.

A corporate website, built in this way, is called a dynamic website because it is "produced" or "generated" via, on the one hand, static HTML templates or models and, on the other hand, structured and unstructured data stored either in a database server or a file server.

We have experienced this approach with very small organizations working in the field of arts and crafts, as well as with research and teaching labs and, finally, also with teachers in the university wanting to possess and manage their website without having to learn technical staff.

Peter Stockinger
Institut National des Langues et Civilisations Orientales (INALCO)
Fondation Maison des Sciences de l'Homme (FMSH)
54, Bd. Raspail – 75006 Paris
email : stockinger@msh-paris.fr
site web : http://www.semionet.fr

This approach of a systematic description of organizational information and communication patterns (also called genre, broadly speaking) can be extended to any kind of dynamic production and management of web-based information or knowledge products and services such as, for instance, for the production and management of interactive encyclopedias, of interactive professional or commercial materials, of interactive pedagogical products, and so on. As we have already mentioned, it seems to have a promising future in almost all concrete domains of applications, such as:

- in e-commerce (building and maintaining of web-based scenarios of B-to-B communication, B-to-C communication, corporate and internal communication, virtual sales points, virtual marketing services, business information watch services, etc.);

- in e-culture ((building and maintaining of web-based scenarios of digital multimedia libraries, virtual exhibitions and art collections, cultural institutions, etc.);

- in e-learning ((building and maintaining of web-based scenarios of virtual campus, virtual learning "institutes", on-line courses, on-line exams, student info services, student registration services, teacher's virtual "authoring" studios, teacher/student communication services, digital library services, ...);

- in e-(web) media (building and maintaining of web-based scenarios of multimedia contents archives, info genres, fictional genres, ...);

- in e-tourism (building and maintaining of web-based scenarios of travel offers, travel preparation services, tourist information products, etc.);

- in e-community (social and community services, administrative information, "direct democracy, ...).

Peter Stockinger
Institut National des Langues et Civilisations Orientales (INALCO)
Fondation Maison des Sciences de l'Homme (FMSH)
54, Bd. Raspail – 75006 Paris
email : stockinger@msh-paris.fr
site web : http://www.semionet.fr

Indeed, it constitutes actually one of the major fields of R&DT and sticks to the convergence of two already central technologies:

- the *XML technologies* for structuring, exchanging and sharing information and

- the *web service technologies* for a user adapted re-use of generic - general or specialized - tools and methods for building and managing not only easily configurable corporate websites but also any kind of information or knowledge product or service.

The place and role of e-semiotics in this concert of technologies, services, and tools is easily identifiable and a highly strategic one : it consists, as we know it already, in the description, in the modelling of *information genres* (of the above quoted stereotypical information and communication patterns that determine, globally, social organizations considered information and knowledge producers and users as well as, locally, its different information or knowledge products and services.

The costs of necessary resources in (conceptual, technological, managerial, and commercial) knowledge management are a serious bottleneck for small and very small social organizations of any type to become present on the web and to exploit the potentialities of the new information and communication technologies. Many small and very small organizations have to externalize not only the development but also the management of their web communication, including the building of their corporate website, their different catalogs and commercial services, their internal information and communication services, and so on.

Peter Stockinger
Institut National des Langues et Civilisations Orientales (INALCO)
Fondation Maison des Sciences de l'Homme (FMSH)
54, Bd. Raspail – 75006 Paris
email : stockinger@msh-paris.fr
site web : http://www.semionet.fr

**7) Final remarks**

Let us summarize, e-semiotics:

- Is a set of conceptual and technical means of the modelling, the description or again the scenario building of digital information and communication services and products,

- Is based on a rich theory and a long experience in the analysis of all kinds of information loaded objects as well as of social situations of communication and information exchanges,

- Works in co-ordination with available technologies (either in form of standards or pieces of software or services and applications).

The first point sticks to the fact that e-semiotics (or semiotics, tout court) is concerned with what is called the conceptual or, again, the cognitive level of information and communication services and products. It uses technology, but it doesn't produce by itself technology. It is much more concerned with enhancing the value and the usability of technology.

The second point sticks to the important, even essential fact that e-semiotics takes advantage of the typical organization patterns of information loaded objects such as technical documents, audiovisual resources, learning products, and so on. This is maybe the most important difference between e-semiotics and other tools and methodologies concerned with the conception and development of information and communication services and products.

Contrarily to most other - functional, object-oriented, process-oriented, etc. - methodologies, it is based mainly on the assumption that the development and real exploitation (i.e., the integration and real use) of digital information and communication products depends crucially on the

Peter Stockinger
Institut National des Langues et Civilisations Orientales (INALCO)
Fondation Maison des Sciences de l'Homme (FMSH)
54, Bd. Raspail – 75006 Paris
email : stockinger@msh-paris.fr
site web : http://www.semionet.fr

"good" understanding and manipulation of their internal, structural organisation - structural organization, sometimes called genre, which they share, at least partially, with "traditional" information and communication products.

The third point sticks to the fact that e-semiotics produces scenarios and models in the field of digital information and communication products and services - not only web-based ones.

As we have already mentioned, e-semiotics is a tool that can be used in different contexts:

- As a means for the conception and development of services and products,
- As a means for evaluating digital information products and services by their users,
- As a means for the production and testing of "models" of a digital corporate communication,
- As a means for the production of "standards", of "models" with the help of which the conceptual part of an information or knowledge management is covered (for instance, the thematic or conceptual organisation of information data),
- As a means for the maintenance and necessary adaptation of existing information and communication products
- Etc.

In this sense, e-semiotics as a conceptual device for building, testing, and renewing information and communication products and services is concerned by the major activities in an information project, such as:

- Information watch
- Information organisation

Peter Stockinger
Institut National des Langues et Civilisations Orientales (INALCO)
Fondation Maison des Sciences de l'Homme (FMSH)
54, Bd. Raspail – 75006 Paris
email : stockinger@msh-paris.fr
site web : http://www.semionet.fr

- Information exploitation
- Information & communication
- Information management

As already mentioned, several times, one of the major R&DT trends goes to the development of libraries of more or less specialized web services in e-commerce, e-culture, e-learning, etc. E-semiotics is concerned here with the definition but also the maintenance and enhancement of the underlying models and scenarios. In my opinion, this constitutes one of the most important - technological, social, and economic challenges for the coming years.

Peter Stockinger
Institut National des Langues et Civilisations Orientales (INALCO)
Fondation Maison des Sciences de l'Homme (FMSH)
54, Bd. Raspail – 75006 Paris
email : stockinger@msh-paris.fr
site web : http://www.semionet.fr